\documentclass{webofc}
\RequirePackage{orcidlink}
\usepackage[varg]{txfonts}   % Web of Conferences font
\usepackage{hyperref}
\usepackage{url}
\hypersetup{colorlinks=true,citecolor=blue,urlcolor=blue,linkcolor=blue}
\begin{document}
% CHEP24-532
\title{Addressing tokens dynamic generation, propagation, storage and
renewal to secure the GlideinWMS pilot based jobs and system}
%
% subtitle is optional
%
%%%\subtitle{Do you have a subtitle?\\ If so, write it here}

\author{
        \firstname{Bruno} \lastname{Moreira Coimbra}\inst{1}\orcidlink{0009-0002-2797-8706}\fnsep\thanks{\email{coimbra@fnal.gov}} \and
        \firstname{Marco} \lastname{Mambelli}\inst{1}\orcidlink{0000-0002-9489-2681}\fnsep\thanks{\email{marcom@fnal.gov}}
}

\institute{
    Fermi National Accelerator Laboratory,
    PO Box 500, Batavia IL 60510-5011
}

\abstract{% 200 words max
GlideinWMS has been one of the first middleware in the WLCG community to transition from X.509 to support also tokens. The first step was to get from the prototype in 2019 to using tokens in production in 2022. This paper will present the challenges introduced by the wider adoption of tokens and the evolution plans for securing the pilot infrastructure of GlideinWMS and supporting the new
requirements.
In the last couple of years, the GlideinWMS team supported the migration of experiments and resources to tokens. Inadequate support in the current infrastructure, more stringent requirements, and the higher spatial and temporal granularity forced GlideinWMS to revisit once more how credentials
are generated, used, and propagated.
The new credential modules have been designed to be used in multiple systems (GlideinWMS, HEPCloud) and use a model where credentials have type, purpose, and different flows.
Credentials are dynamically generated in order to customize the duration and limit the scope to the targeted resource. This allows to enforce the least privilege principle. Finally, we also considered adding credential storage, renewal, and invalidation mechanisms within the GlideinWMS infrastructure to better serve the experiments’ needs.
}
\maketitle
\section{Introduction}
\label{intro}
%placeholder to remove
GlideinWMS streamlines resource provisioning in distributed High Throughput Computing (HTC) environments, efficiently scouting for resources, tailoring worker nodes, and ensuring robust monitoring and auditing. By leveraging HTCondor, it optimizes job execution while maintaining system homogeneity. Secure authentication and credential management are essential to preserve the integrity of distributed systems. Over time, traditional X.509 certificate-based security has given way to token-based authentication, enhancing both scalability and usability. This paper delves into the intricacies of token management and examines how GlideinWMS compares to other security frameworks in distributed computing.

\section{GlideinWMS}
\begin{figure}[htpb]
    \centering
    \includegraphics[width=8cm,clip]{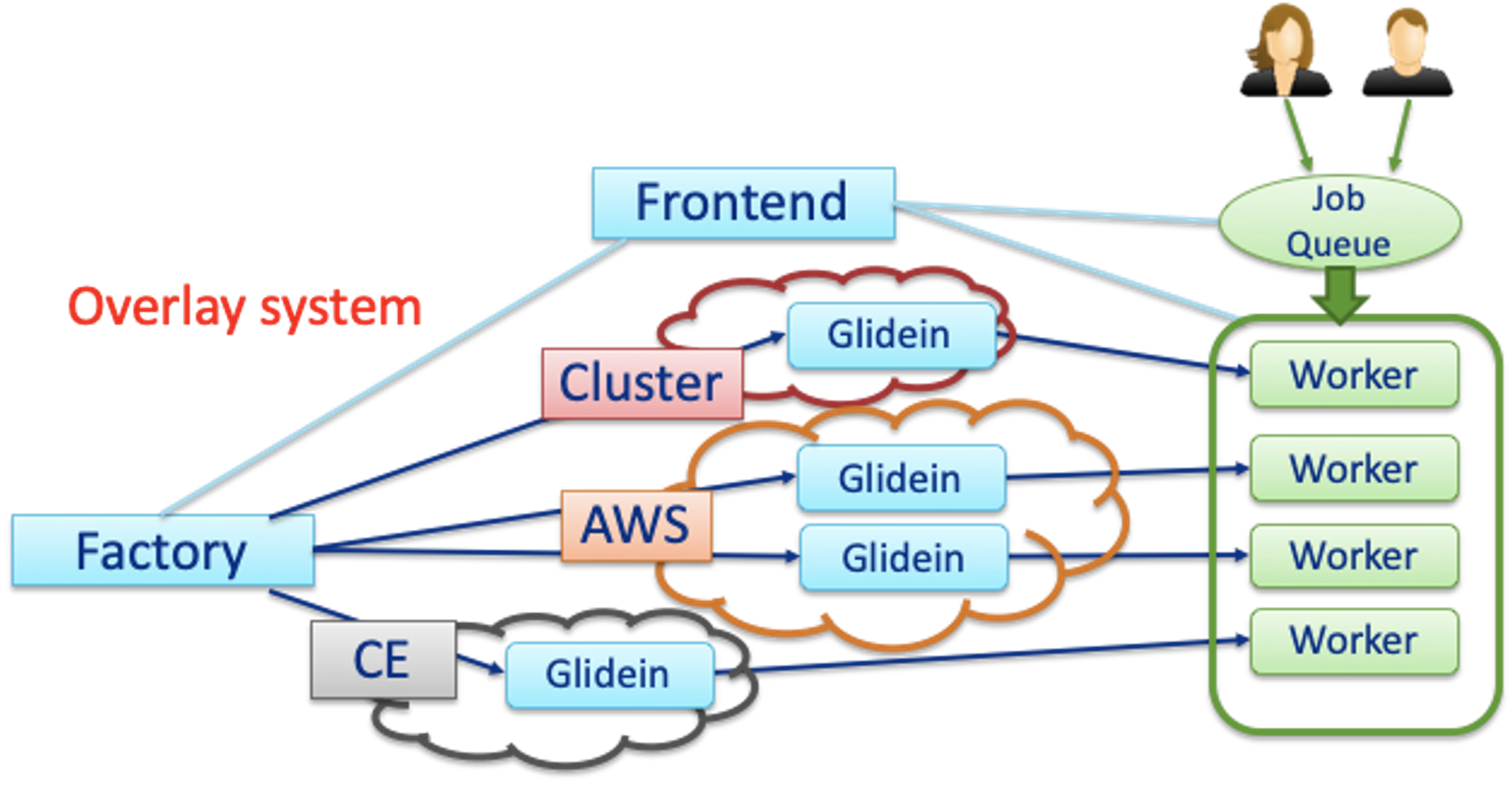}
    \caption{GlideinWMS system. GlideinWMS components are in blue, the User Pool is in Green, and the computing resources are in other colors.}
    \label{fig-1}
\end{figure}

GlideinWMS~\cite{glideinwms,gwms-sw} is a pilot-based Workload Management System (WMS) that provisions computing resources in a distributed environment. Users can request customized elastic HTCondor Software Suite (HTCSS)\cite{htcondor} clusters and User Pools, as shown in figure~\ref{fig-1}, where they run their computations. GlideinWMS sends Glideins, also called pilot jobs, to various computing resources, distinguishing them from user jobs. It has been and is used at scale in production by many collaborations, including the Compact Muon Solenoid (CMS) experiment, many Fermilab experiments, and the OSG for over 10 years. The Virtual Organization (VO), the computing model abstraction of these collaborations, is used interchangeably with GlideinWMS. Most scientists interact with tools or portals like CRAB, JobSub, or OSG-Connect provided by the scientific collaborations instead of GlideinWMS or its clusters.

The Glidein, or pilot job, configures computing resources to run user jobs by detecting node resources, installing common tools, and managing credentials. It also reports system status and joins the User Pool for job execution.

The Factory and clients like the VO Frontend or HEPCloud’s Decision Engine form the GlideinWMS system. This paper considers a setup with one Factory, one Frontend, and their Glideins, though real deployments may include multiple clients and Factories for redundancy.

The Factory submits Glideins to Compute Entrypoints (CEs), handling resource access, supported VOs, authentication, and throttling. It monitors Glideins, caches credentials, and provides a secure mailbox for client requests and updates.

Clients, including the Frontend, track user requests and available Glideins, adjusting resource allocations via heuristics to ensure efficiency while adhering to policies.

Typically operated by VOs, the Frontend manages key credentials, such as Factory authentication, Glidein submission, user pool integration, and VO-level services (monitoring, databases, storage).

\section{Security and Credentials in GlideinWMS}
\label{sec-cred}
Security in GlideinWMS relies on multiple credential types used for different authentication needs, aka purposes:

\begin{itemize}
    \item \textbf{Pilot Submission Credentials (P-CRED):} Provide access to CEs for Glideins.
    \item \textbf{VO Service Credentials (S-CRED):} Access secured VO resources such as databases.
    \item \textbf{CE Credentials (CE-CRED):} Used for monitoring and auditing on the CE.
    \item \textbf{HTCondor Cluster Credentials (C-CRED):} Issued by the HTC Central Manager to join the User Pool.
    \item \textbf{Job Credentials (J-CRED):} Job specific secure storage access tokens.
    \item \textbf{Framework Credentials (F-CRED):} Framework tokens used to authenticate Factories and Frontends with each other.
\end{itemize}

Each of these credentials can use various authentication methods, including SciTokens\cite{scitokens}, IDTokens, and traditional SSH keys.

\subsection{Token Authentication Mechanisms}
\label{sec-cred-token}
The deprecation of X.509 proxies, no longer maintained by the software libraries, motivated the GlideinWMS transition to identity and access tokens\cite{gwms-tokens-chep23}. With the phasing out of these proxies, GlideinWMS had to adapt to a new security model that could handle multiple credential types with different functionalities, such as identity tokens, access tokens, JWT-based tokens, and other certificates.

The transition to token-based authentication involves changes for all interactions.
For the Framework and the HTCondor virtual cluster, this involves:
IDTokens replacing legacy GSI authentication, centralized authentication using HTCondor Security, and issuance, renewal, and validation of per-resource tokens.

The security model and the history of the transition are described in more detail in our previous paper \cite{gwms-tokens-chep23}.
This one will focus on the handling of tokens, the dynamic credential generation, and the software development that followed that initial work.

\section{Credentials Module}
\label{cred-module}
The implementation of token support in GlideinWMS introduced complexity to parts of the code and highlighted challenges to support multiple credential types. Adding a new credential type required editing many functions, often adding the type to lists, creating new conditions, or creating new conditional statements. This exponentially increased code complexity by introducing new branches of execution in already long functions.

To address these challenges, a new credentials module was created. It defines hierarchical credential classes, moving complexity from business logic to specific class implementations. Developers can use a generic credential base class, handling credentials throughout the code without considering specific types. When implementing a new type, developers can focus on basic operations without worrying about code impact. The module also defines credential purposes, improving workflow decision-making, and introducing new functionalities.

More information about the Credential base class is in the next section.

\subsection{Credential Base Class}
The Credential base class is an abstract class that defines the basic functionalities expected to be implemented by every concrete credential type. This class holds the credential payload, an attribute that stores the underlying credential object to be used as the source of truth for all other credential attributes.

The only value we store for an object of the Credential class is the string representation of a credential stored in a private variable. All other attributes of the credential are implemented as class properties and derived from the stored credential string when they are accessed. This model not only simplifies the implementation of child credential classes but also prevents inconsistencies between the credential data represented by its string and the attributes of the credential object.

The underlying credential object of a concrete credential type implementation of the Credential class is accessed through a private payload property. This property implementation returns the decoded string of the credential, using the defined decode function for that credential type. With this mechanism in place, specific credential attributes can easily be implemented by calling simple functions from the payload variable.

The code below shows simplified implementations of the Credential base class and a child Token class. Note that this implementation intends to exemplify the techniques discussed in this section and omits several implementation details.

\begin{verbatim}
class Credential(ABC, Generic[T]):
    self._string = None

    @property
    def string(self) -> Optional[bytes]:
        return self._string

    @property
    def _payload(self) -> Optional[T]:
        return self.decode(self.string) if self.string \
        else None

    @staticmethod
    @abstractmethod
    def decode(string: Union[str, bytes]) -> T:
        pass

class Token(Credential[Mapping]):
    @property
    def subject(self) -> Optional[str]:
        return self._payload.get("sub", None) if self._payload \
        else None

    @property
    def scope(self) -> Optional[str]:
        return self._payload.get("scope", None) if self._payload \
        else None

    @staticmethod
    def decode(string: Union[str, bytes]) -> Mapping:
        if isinstance(string, bytes):
            string = string.decode()
        return jwt.decode(string.strip())
\end{verbatim}

\subsection{Credential Types}
Using the Credential base class and the mechanisms discussed in the previous section, several concrete implementations of specific credential types were developed. Each of these credential types inherits from the Credential base class and implements all the necessary abstract methods and the properties unique to its type. Moreover, concrete credential types can be hierarchical, allowing them to have their own child credential types. In the current version of the Credentials module, there’s an example of a hierarchical credential type: the Token credential type, which implements a basic JSON Web Token (JWT). This credential type has two child types: HTCondor IDTOKEN and SciToken. Both child types have the same methods but define their own naming standards and file extensions.

A diagram illustrating the current hierarchy of credential types available in the module is presented in figure \ref{fig:cred-hierarchy}.

\subsection{Credential Pairs}
Some authentication methods combine public and private credentials (asymmetric authentication). The GlideinWMS Credentials module introduces the CredentialPair base class to manage these scenarios. It inherits attributes and methods from the regular Credential class but includes a private credential attribute that holds another instance of the Credential class. Credential pairs are treated as simple credentials in most GlideinWMS code.

Credential pairs inherit from both the CredentialPair base class and their underlying credential type. For example, the X509Pair class inherits from CredentialPair and X509Cert classes. This polymorphism allows credentials to be treated with varying levels of complexity in different GlideinWMS workflows.

\begin{figure}[h]
    \centering
    \begin{minipage}{0.48\linewidth}
        \raggedright
        \includegraphics[width=\linewidth]{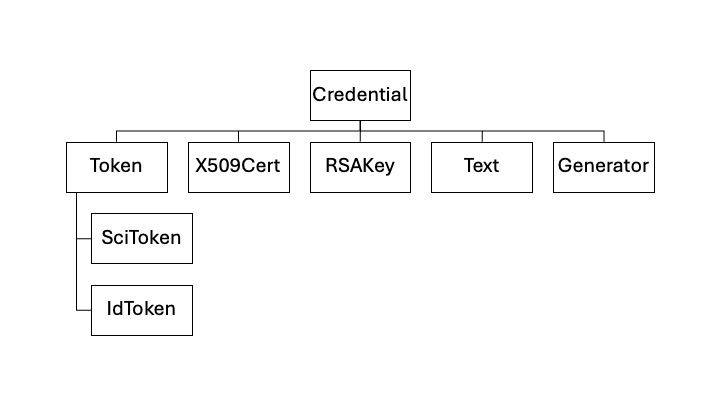}
        \caption{Credentials Types Hierarchy}
        \label{fig:cred-hierarchy}
    \end{minipage}
    \hfill
    \begin{minipage}{0.48\linewidth}
        \raggedleft
        \includegraphics[width=\linewidth]{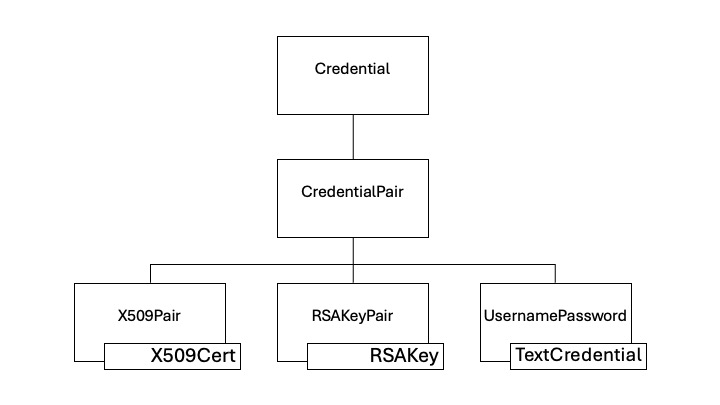}
        \caption{Credentials Pairs Hierarchy}
        \label{fig:cred-pair-hierarchy}
    \end{minipage}
\end{figure}

A diagram illustrating the current credential pairs included in the Credentials module is presented in figure \ref{fig:cred-pair-hierarchy}.

\subsection{Credential Purposes}
The credentials module defines three credential purposes that can be assigned to credentials. These purposes are directly related to the purposes discussed in section \ref{sec-cred} that GlideinWMS handles. They are:

\begin{itemize}
    \item \textbf{request} (P-CRED): Used to authenticate with the compute element (CE) of a site.
    \item \textbf{payload} (S-CRED): Payload credentials are sent along with the Glidein but not used in the GlideinWMS workflow. They are often used by operators to authenticate with services or resources from worker nodes.
    \item \textbf{callback} (C-CRED): Used by the startd spawned by a Glidein on a worker node to authenticate back with an HTCondor collector that manages the user pool.
\end{itemize}

\subsection{Credential Generators}
When dealing with minimal-privileged credentials like tokens, it’s often necessary to use different credentials for clients and hosts. This can be managed with pre-generated credentials for a small number of cases, but becomes increasingly challenging when dealing with hundreds of sites. Additionally, credentials can have short lifespans and require frequent renewal. To address these issues, the Credentials module offers credential generators.

The CredentialGenerator class leverages the GlideinWMS Generator Framework to load Python scripts capable of generating specific credential types based on provided arguments. This feature enables the generation of credentials at runtime on a per-site basis, simplifying configuration files. Moreover, since they are dynamically created, the risk of unintended credential exposure is substantially reduced.

\subsection{Security Parameters}
The Credential module introduces security parameters for authentication with sites. These parameters pass additional context when requesting computing resources, such as the resource type or account. Previously, these qualifiers were available in GlideinWMS as credentials. The redesigned parameters are now declared independently and considered during authentication for any selected credential on a site. Like credentials, security parameters are defined by the Parameter base class and have data types (Integer, String, or Expression). Each type inherits from the base class and implements unique methods.

\subsection{Parameter Generators}
As their name suggests, generator parameters also leverage the GlideinWMS Generator Framework to dynamically generate their values. They have access to the same runtime arguments as credential generators. Generated parameters can be used to dynamically add context to an authentication identity based on the site resources being requested from or other criteria defined by the generator itself.

\section{GlideinWMS Generators Framework}
The GlideinWMS Generators Framework is a set of tools to help users generate data at runtime to be used at different times in the GlideinWMS execution workflow. The framework consists of a set of built-in generators ready to be used straight from the GlideinWMS Factory and Frontend configuration files, and a Python module that provides the tools for users to write their own generators.

\subsection{Usage}
Generators are currently supported by security credentials and parameters. Users can directly use generators from the GlideinWMS configuration files. When declaring credentials or parameters, users should set their type to “generator” and specify the generator name in the “absfname” property for credentials or the “value” property for parameters. Generators also require additional settings passed in the "context" property as a dictionary. The content of these dictionaries varies by generator, but at a minimum, they must contain the “type” key, which specifies the type of content generated.

Next, two configuration examples are presented. One demonstrates how to set up a credentials generator, while the other illustrates how to configure a parameter generator.

\begin{verbatim}
<credential
    absfname="RoundRobinGenerator" purpose="payload"
    security_class="frontend" trust_domain="grid"
    context="{'items': ['str1', 'str2', 'str3'], 'type': 'text'}"
    type="generator"
/>

<parameter
    name="VMId" value="RoundRobinGenerator"
    context="{'items': ['vm1', 'vm2', 'vm3'], 'type': 'string'}"
    type="generator"
/>
\end{verbatim}

\subsection{Custom Generators}
Users can create custom generators if the built-in ones do not meet their specific needs. A custom generator is a Python module that exports a subclass of the Generator base class. This subclass implements the "generate" method to return the generated content. The example below shows a custom generator that returns a random element of a list passed in the "items" key of the context dictionary.

\begin{verbatim}
import random
from typing import Any
from glideinwms.lib.generators import export_generator, Generator, GeneratorError

class RandomGenerator(Generator[Any]):
    """Random generator"""
    def generate(self, **kwargs) -> Any:
        items = self.context.get("items", [])
        if not items:
            raise GeneratorError("No items provided for generation")
        return random.choice(items)

export_generator(RandomGenerator)
\end{verbatim}

There are three key points in this example. First, the class is exported using the "export\_generator" method from the Generators module. This method adds the class to the GlideinWMS environment, enabling its runtime import. Second, the "context" dictionary is accessed without being explicitly declared in the RandomGenerator class. This is because it is declared in the Generator parent class. Third, the generate method accepts arbitrary keyword arguments. These arguments are used to access runtime arguments provided by the GlideinWMS system. A list of the currently provided arguments is available in the GlideinWMS documentation \cite{gwms-doc}.

\subsection{Legacy Generators}
Previous GlideinWMS versions supported credential generators using the Callout API, which provided similar functionalities but lacked flexibility in generating security parameters and integrating with the configuration. Users are encouraged to rewrite callout-based generators using the new framework, but can also use the built-in LegacyGenerator, an adapter for callout scripts that can be configured directly from the Frontend configuration. Here’s an example of adding a LegacyGenerator to the front-end configuration.

\begin{verbatim}
<credential
    absfname="LegacyGenerator" purpose="payload"
    security_class="frontend" trust_domain="grid"
    context="{
        'callout': 'example_callout.py',
        'type': 'scitoken',
        'kwargs': {'param1': 'value1', 'param2': 'value2'}
    }"
    type="generator"
/>
\end{verbatim}

\section{Conclusions}
\label{sec-conc}
% placeholder to replace
The GlideinWMS security framework has evolved to incorporate token-based authentication, significantly improving system security and flexibility. By implementing structured credential classes, automated token management, and renewal mechanisms, we reduce technical debt in the code and enhance authentication reliability while minimizing administrative overhead. 

GlideinWMS provides a seamless and scalable approach to token-based authentication in HTC environments. Future work will focus on refining credential refresh mechanisms and integrating additional security enhancements.

\section{Acknowledgments}
\label{ack}
The authors’ work was performed using the resources of the Fermi National Accelerator Laboratory (Fermilab), a U.S. Department of Energy, Office of Science, HEP User Facility. Fermilab is managed by Fermi Forward Discovery Group, LLC, acting under Contract No. 89243024CSC000002. 

% BibTeX or Biber users please use (the style is already called in the class, ensure that the "woc.bst" style is in your local directory)
\bibliography{gwms-hepcloud}
% Replace "your_bib_file" with the actual name of your .bib file
%
% Non-BibTeX users please use
%
%\begin{thebibliography}{}
%
% and use \bibitem to create references.
%
%\bibitem{RefJ}
% Format for Journal Reference
%Journal Author, Article title. Journal \textbf{Volume}, page numbers (year). %\url{https://doi.org/Article-DOI-number}
% Format for books
%\bibitem{RefB}
%Book Author, \textit{Book title} (Publisher, place, year) page numbers
% etc
%\end{thebibliography}

\end{document}